\documentclass[12pt,a4paper]{article}
\usepackage[T2A]{fontenc}
\usepackage[utf8]{inputenc}
\usepackage[english]{babel}
\usepackage{amssymb}
\usepackage{amsmath}
\usepackage{graphicx}
\usepackage[small,sc,center]{titlesec}
\usepackage{indentfirst}
\usepackage{siunitx}
\usepackage[labelsep=period]{caption}
\usepackage{caption}

\newcommand{\nc}{\newcommand}
\nc{\Teff}{T_\mathrm{eff}}
\nc{\tev}{t_\mathrm{ev}}
\begin{document}

\begin{center}
	\textbf{Evolutionary status and fundamental parameters of the Cepheid V1033~Cyg}
	
	\vskip 3mm
	\textbf{Yu. A. Fadeyev\footnote{E--mail: fadeyev@inasan.ru}}
	
	\textit{Institute of Astronomy, Russian Academy of Sciences,
		Pyatnitskaya ul. 48, Moscow, 119017 Russia} \\
	
	Received June 15, 2020; revised June 17, 2020; accepted June 26, 2020
\end{center}

\textbf{Abstract} ---
On the basis of consistent stellar evolution and nonlinear stellar pulsation
calculations the Cepheid V1033~Cyg is shown to be the post--main sequence star
at the first crossing of the instability strip during gravitational contraction
of the helium core.
The observed light variability of V1033~Cyg is due to radial oscillations
in the fundamental mode.
The best agreement (within one percent) between recent observations and
the theoretical estimate of the period change rate was obtained for the
evolutionary sequence with stellar mass $M=6.3M_\odot$ and helium and heavier
element fractional abundances $Y=0.28$ and $Z=0.022$, respectively.
The age of the star, the luminosity, the radius, the effective temperature and
the surface gravity are $\tev=5.84\times 10^7$~yr, $L=2009L_\odot$,
$R=45.6R_\odot$, $\Teff=5726$~K, $\log g=1.92$.


\newpage
\section*{introduction}

According to the General catalogue of variable stars (Samus' et al. 2017)
the pulsating variable V1033~Cyg with period of light variations $\Pi=4.9375$~day
belongs to $\delta$~Cep type stars.
The most remarkable feature of V1033~Cyg is the enormously high rate of period
increase.
The study of the secular period change shows that the $O-C$ diagram spanning
last 117 yr can be described with a high accuracy by the quadratic dependence,
whereas the period change rate is $\dot\Pi=18.19$~s/yr with a r.m.s. error
$\sigma\approx 0.08$~s/yr (Berdnikov et al. 2019).
This value of $\dot\Pi$ is several dozen times larger than that of most
core helium burning Cepheids (Fadeyev 2013, 2014).
V1033~Cyg seems to be in the relatively short evolutionary stage of gravitational
contraction of the helium core during which the star crosses the
Hertzsprung--Russel (HR) diagram in the thermal time scale.
This assumption is confirmed by detection of ${}^7$Li lines indicating
that the star is in the stage preceding the red giant
(Luck and Lambert 2011).
Thus, V1033~Cyg seems to be the second Cepheid known as a post--main sequence
star crossing the instability strip for the first time.
Earlier the evolutionary status of the Cepheid at the first crossing
has been established for $\alpha$~UMi (Turner et al. 2005; Fadeyev 2015a).

Owing to the small error of the observational estimate of $\dot\Pi$
the Cepheid V1033 becomes a very appropriate object for determination of its
fundamental parameters (i.e. the stellar age $\tev$, the mass $M$,
the luminosity $L$ and the radius $R$) by means of comparison
of theoretically computed $\Pi$ and $\dot\Pi$ with observations.
The method is based on consistent calculations of stellar evolution and
nonlinear stellar pulsations where selected models of evolutionary sequences
are used as initial conditions in solution of the Cauchy problem for the
equations of radiation hydrodynamics and time--dependent convection
describing radial stellar pulsations.
Earlier this method was employed for determination of the fundamental parameters
of the Cepheids $\alpha$~UMi (Fadeyev 2015a) and SZ~Tau (2015b) as well as
of groups of long--period (Fadeyev 2018) and short--period Cepheids (Fadeyev 2020).
Efficiency of the method is confirmed by a good agreement (within several percent)
of the results of theoretical computations with Baade--Wesselink radius
measurements for long--period Cepheids RS~Pup, GY~Sge and SV~Vul.

The goal of the present study is to confirm the evolutionary status of V1033~Cyg
as a Cepheid crossing the instability strip for the first time as well as
to determine its fundamental parameters by methods of the theory of stellar
evolution and the theory of stellar pulsation.
Evolutionary computations were done with the program MESA version 12115
(Paxton et al. 2018).
Calculations of evolutionary sequences and nonlinear stellar pulsations are
discussed in our preceding paper (Fadeyev 2020).
Initial conditions necessary for solution of the equations of hydrodynamics
were obtained from selected models of evolutionary sequences with initial masses
$5.7M_\odot\le M\le 7.4M_\odot$ and initial fractional abundance of helium
$Y=0.28$.
Bearing in mind uncertainties in Cepheid metallicities due to the existence
of the galactic abundance gradient
(Luck and Lambert 2011; Luck et al. 2011; Lemasle et al. 2018)
the stellar evolution and the stellar pulsation calculations were done for
the standard metallicity $Z=0.02$ as well as for $Z=0.018$ and $Z=0.022$.

\section*{results of calculations}

Evolutionary sequences were computed from the initial main sequence up to
the stage preceding the evolution along the red giant branch on the HR diagram.
Effects of mass loss due to the stellar wind in main sequence stars with masses
$M\le 7.4M_\odot$ remain negligible since the star loses less than
$10^{-4}$ of its total mass.
Moreover, in the outer layers with the temperature $T\lesssim 5\times 10^6$~K
and the radius $r\ge 0.1R$, where pulsation motions are perceptible, abundances of
chemical elements do not undergo evolutionary changes.
Therefore, without significant loss of accuracy we can assume that the masses
of Cepheid hydrodynamic models as well as abundances of chemical
elements are the same as on the initial main sequence.

During the first crossing of the instability strip the pulsation period
monotonically increases, evolution of stars with masses $M\le 7M_\odot$
being accompanied by switch of oscillations from the first overtone to the
fundamental mode.
To determine with sufficient accuracy the star age corresponding to pulsation mode
switching we computed for each evolutionary sequence $\approx 15$ hydrodynamic models
uniformly distributed with respect to $\tev$ between edges of the instability strip.
According to approach employed in our preceding study (Fadeyev 2019) we
assumed that the star age at the pulsation mode switch is the mean value
of the ages of two adjacent hydrodynamic models pulsating in different modes.
Within the interval with continuous change of the pulsation period
the dependence $\Pi(\tev)$ was calculated using a least-squares polynomial fit.
For polynomials of the third order the r.m.s. error of approximation does not
exceed 0.1 percent.

\begin{figure}
\centerline{\includegraphics{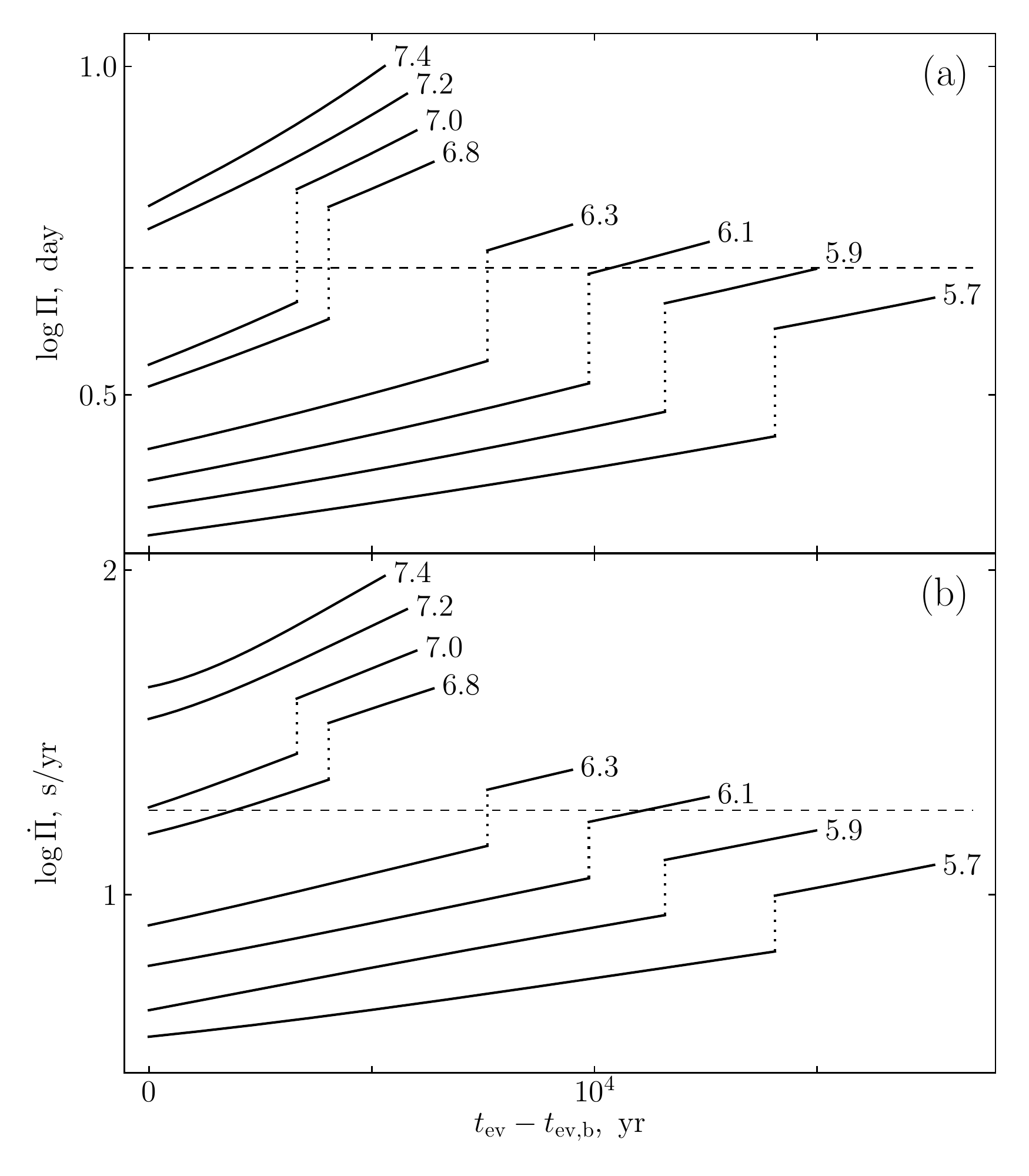}}
\caption{Temporal variations of the pulsation period (a) and of the rate of period change (b)
         in radially pulsating Cepheids with $Z=0.02$ during the first crossing of the
         instability strip.
         The stellar masses $M$ are indicated at the tracks.
         Dotted lines correspond to pulsation mode switching.
         Horizontal dashed lines show the observational estimates of the period
         $\Pi=4.9403$ day and of the period change rate $\dot\Pi=18.19$\ s/yr
         (Berdnikov et al. 2019).}
\label{fig1}
\end{figure}

Results of consistent stellar evolution and nonlinear stellar pulsation computations
are illustrated in Fig.~\ref{fig1} where the plots of evolutionary variations of $\Pi$
and $\dot\Pi$ are shown for Cepheids with metal abundance $Z=0.02$.
For the sake of graphical convenience the horizontal coordinate is set to zero when
the star of the age $t_\mathrm{ev,b}$ crosses the blue edge of the instability strip.
The final point of each plot represents decaying pulsations at the red edge
of the instability strip.
The duration of the first crossing of the instability strip lasts between
$\approx 5.3\times 10^3$ yr for $M=7.4M_\odot$ and $\approx 1.8\times 10^4$ yr
for $M=5.7M_\odot$.
The horizontal dashed lines in Figs.~\ref{fig1}a and \ref{fig1}b
are shown for comparison of the results of computations with observations
and correspond to the period $\Pi=4.9403$~yr and the rate of period change
$\dot\Pi=18.19$~s/yr (Berdnikov 2019).

The abrupt period change shown in plots by dotted lines is due to pulsation
mode switching from the first overtone to the fundamental mode.
As can be seen in Fig.~\ref{fig1}a, the pulsation mode switching significantly
constrains the mass interval of stars pulsating with period $\Pi\approx 4.94$ day.
In particular, this condition is fulfilled for the fundamental mode pulsators
of evolutionary sequences with masses $5.9M_\odot\lesssim M\lesssim 6.1M_\odot$.

\begin{figure}
\centerline{\includegraphics{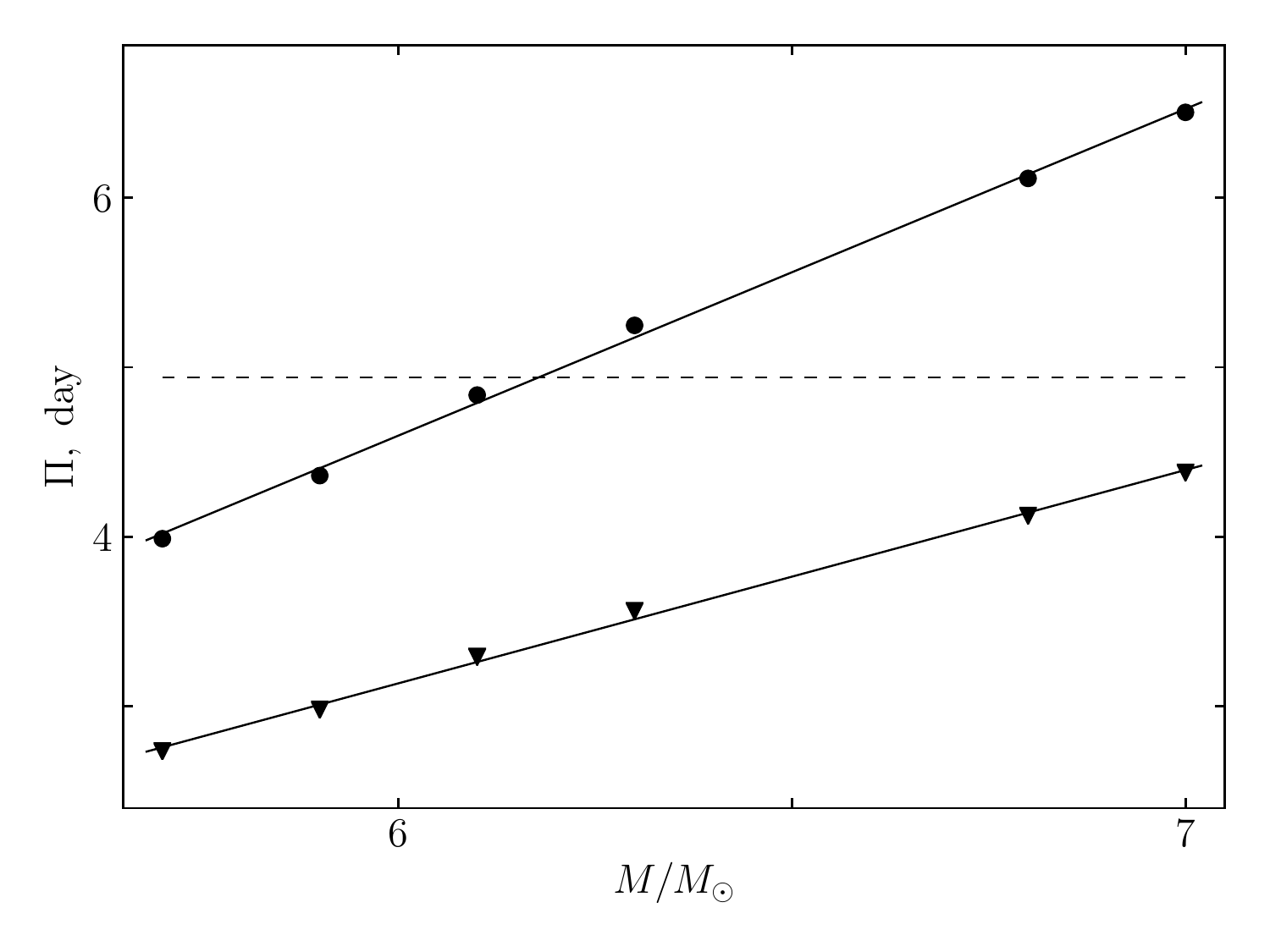}}
\caption{Periods of the fundamental mode $\Pi_0$ (filled circles) and of the first overtone
         $\Pi_1$ (filled triangles) at the point of mode switching as a function of stellar
         mass $M$ for evolutionary sequences with metallicity $Z=0.02$.
         The horizontal dashed line shows the period $\Pi=4.9403$ day.}
\label{fig2}
\end{figure}

In Cepheids at the first crossing of the instability strip the periods of the 
fundamental mode $\Pi_0$ and of the first overtone $\Pi_1$ at the point of
pulsation mode switching monotonically increase with increasing stellar mass.
For evolutionary sequences with initial masses ranged within $5.7M_\odot\le M\le 7.0M_\odot$
the increase of the period is described by relations
\begin{equation}
\label{pmsw0}
\Pi_0 = -6.962 + 1.926 M/M_\odot ,
\end{equation}
\begin{equation}
\label{pmsw1}
\Pi_1 = -4.413 + 1.258 M/M_\odot ,
\end{equation}
where $\Pi_0$ and $\Pi_1$ are expressed in days.
Dependencies (\ref{pmsw0}) and (\ref{pmsw1}) represent the linear approximation
of $\Pi_0$ and $\Pi_1$ at the points of mode switching (see Fig.~\ref{fig2})
and were determined from hydrodynamic computations.

Attempts to compute the hydrodynamic model of the Cepheid V1033~Cyg pulsating
in the first overtone have failed.
This is due to the two following reasons.
First, radial oscillations in the first overtone are excited in stars with
masses $M\le 7M_\odot$ and the upper limit of the period is
$\Pi\approx 4.4$ day.
Second, if we succeed to compute the hydrodynamic model for the first overtone pulsator
with mass $M>7M_\odot$ (for example, via variations of stellar chemical composition)
then, as can be seen in Fig.~\ref{fig1}b, the growth of the period change rate with
increasing stellar mass inevitably leads to contradiction with observations.

\begin{figure}
\centerline{\includegraphics{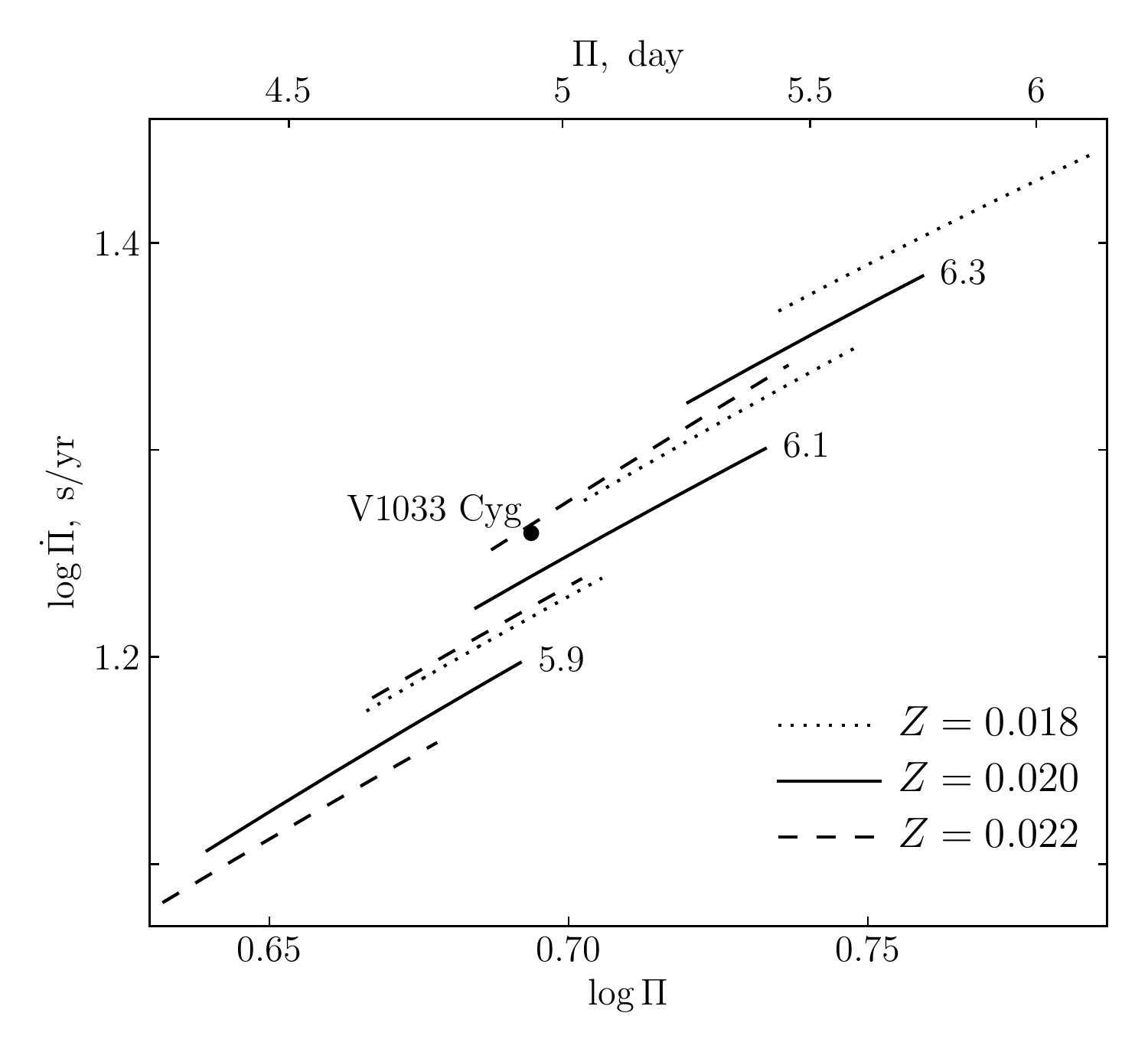}}
\caption{The rate of period change $\dot\Pi$ as a function of period $\Pi$ for
         Cepheid models pulsating in the fundamental mode under different assumptions
         on metallicuty $Z$.
         The numbers at the plots indicate the stellar mass $M$.
         The filled circle shows the Cepheid V1033~Cyg (Berdnikov et al.2019).}
\label{fig3}
\end{figure}

To obtain more accurate estimates of the fundamental parameters of V1033~Cyg
determined from hydrodynamic models of the fundamental mode pulsators
we computed additional evolutionary sequences of the stars with masses
$M=5.9M_\odot$, $6.1M_\odot$ and $6.3M_\odot$ for metallicities $Z=0.018$ and
$Z=0.022$.
Results of calculations are presented in the diagram period--rate of period change
shown in Fig.~\ref{fig3}.
The plots demonstrate evolutionary changes of the fundamental mode period and
of the rate of period increase between the point of mode switching and the red
edge of the instability strip.
As can be seen in Fig.~\ref{fig3}, increase of metallicity $Z$ for a fixed value
of the stellar mass $M$ is accompanied by displacement of the plot to smaller $\dot\Pi$.

As a criterion of agreement between the theory and observations we will use
the quantity $\delta = \vert\dot\Pi/\dot\Pi_\star - 1\vert$,
where $\dot\Pi_\star=18.19$ is the observational estimate of the
period change rate and $\dot\Pi$ is the theoretical estimate corresponding to
the pulsation period of the hydrodynamic model $\Pi=4.9403$ day.
The best agreement with observations ($\delta\lesssim 10^{-2}$)
in the diagram period -- rate of period change is obtained for the
evolutionary sequence $Z=0.022$, $M=6.3M_\odot$.
The difference between theoretical and observational estimates of $\dot\Pi$
increases with decreasing $Z$ and is $\delta\approx 0.05$ for $Z=0.020$
and $\delta\approx 0.1$ for $Z=0.018$.

Fundamental parameters of the Cepheid V1033~Cyg are listed in Table~\ref{tabl1}.
The mass $M$ and metallicity $Z$ given in the first and in the second columns
correspond to the evolutionary sequence with minimum deviation $\delta$.
The evolutionary time $\tev$ in the third column is the age of the
hydrodynamic model pulsating with period $4.9403$~day.
The luminosity $L$, the radius $R$, the effective temperature $\Teff$ and
the surface gravity $g$ were evaluated by interpolation of the results
of evolutionary calculations for the star age $\tev$.
Theoretical period change rates $\dot\Pi$ given in the last column
were calculated using the approximate nonlinear dependences $\dot\Pi(\tev)$.

\section*{conclusions}

Results of computations presented in this paper allow us to confidently conclude
that the Cepheid V1033~Cyg is the post--main sequence star in the evolution
stage of gravitational contraction of the helium core.
The star crosses the instability strip for the first time and moves across
the HR diagram to the region of red giants.
Analysis of our hydrodynamic models confirms assumption on
fundamental mode pulsations of V1033~Cyg proposed by
Udovichenko et al. (2019) on the basis of photometric measurements.
Owing to insignificant role of mass loss in main sequence evolution of
intermediate--mass stars we obtained the fairly accurate theoretical
estimates of the fundamental parameters.
Due to existing uncertainties in metallicities ($0.018\le Z\le 0.022$)
the theoretical estimates of stellar mass range within
$5.9M_\odot\lesssim M\lesssim 6.3M_\odot$.
It should also be noted that our theoretical estimates of surface gravity
($1.91\lesssim\log g\lesssim 1.92$)
are in satisfactory agreement with $\log g = 2.10$ obtained for V1033~Cyg
from calculations of stellar atmosphere models (Martin et al. 2015).

\newpage
\section*{references}

\begin{enumerate}

\item L.N. Berdnikov, E.N. Pastukhova, V.V. Kovtyukh, B. Lemasle, A.Yu. Kniazev, I.A. Usenko,
      G. Bono, E. Grebel, G. Hajdu, S.V. Zhuiko, S.N. Udovichenko, and L.E. Keir,
      Astron. Lett. \textbf{45}, 227 (2019).

\item Yu.A. Fadeyev, Astron. Lett. \textbf{39}, 746 (2013).

\item Yu.A. Fadeyev, Astron.Lett. \textbf{40}, 301 (2014).

\item Yu.A. Fadeyev, MNRAS \textbf{449}, 1011 (2015a).

\item Yu.A. Fadeyev, Astron. Lett. \textbf{41}, 640 (2015b).

\item Yu.A. Fadeyev, Astron. Lett. \textbf{44}, 782 (2018).

\item Yu.A. Fadeyev, Astron. Lett. \textbf{45}, 353 (2019).

\item Yu.A. Fadeyev, Astron. Lett. \textbf{46}, in press (2020).

\item B. Lemasle, G. Hajdu, V. Kovtyukh, L. Inno, E.K. Grebel, M. Catelan,
      G. Bono, P. Fran\c cois, A. Kniazev, R. da Silva, and J. Storm,
      Astron. Astrophys. \textbf{618}, A160 (2018).

\item R.E. Luck and D.L. Lambert,  Astron. J. \textbf{142}, 136 (2011).

\item R.E. Luck, S.M. Andrievsky, V.V. Kovtyukh, W. Gieren, W. and D. Graczyk,
      Astron. J. \textbf{142}, 51 (2011).

\item R.P. Martin, S.M. Andrievsky, V.V. Kovtyukh, S.A. Korotin, I.A. Yegorova, I. Saviane,
      MNRAS \textbf{449}, 4071 (2015).

\item B. Paxton, J. Schwab,  E.B. Bauer, L. Bildsten, S. Blinnikov,
      P. Duffell, R. Farmer,  J.A. Goldberg, P. Marchant, E. Sorokina, A. Thoul,
      R.H.D. Townsend, and F.X. Timmes,
      Astropys. J. Suppl. Ser. \textbf{234}, 34 (2018).

\item N.N. Samus', E.V. Kazarovets, O.V. Durlevich, N.N. Kireeva, and E.N. Pastukhova,
      Astron. Rep. \textbf{61}, 80 (2017).

\item D.G. Turner, J. Savoy, J. Derrah, M. Abdel--Sabour Abdel--Latif, and L.N. Berdnikov,
      Publ. Astron. Soc. Pacific \textbf{117}, 207 (2005).

\item S.N. Udovichenko, V.V. Kovtyukh, and L.E. Keir,
       Odessa Astron. Publ. \textbf{32}, 83 (2019).

\end{enumerate}

\newpage
\begin{table}
\caption{Fundamental parameters of the Cepheid V1033~Cyg}
\label{tabl1}
\begin{center}
 \begin{tabular}{cccccccc}
  \hline
 $M/M_\odot$ & $Z$ & $t_\mathrm{ev},\ 10^6$ лет & $L/L_\odot$ & $R/R_\odot$ & $T_\mathrm{eff}$, K & $\log g$ & $\dot\Pi$,~s/yr\\
  \hline
 5.9 & 0.018 &  65.645 &  1648 &  44.6 & 5510 & 1.910 & 16.57 \\
 6.1 & 0.020 &  62.005 &  1802 &  45.1 & 5604 & 1.915 & 17.34 \\
 6.3 & 0.022 &  58.431 &  2009 &  45.6 & 5726 & 1.919 & 18.36 \\
  \hline          
 \end{tabular}
\end{center}
\end{table}
\clearpage

\newpage
\clearpage

\newpage
\clearpage

\newpage
\clearpage

\end{document}